\documentclass[pre%
,preprint%
,draft%
,superscriptaddress%
,floats
,aps%
]{revtex4}
\usepackage{amsmath}%

\begin{document}
\title{Optimizing at the Ergodic Edge} 
\date{\today}
\author{Stefan Boettcher and Martin Frank}
\affiliation{Physics Department, Emory University, Atlanta, Georgia
30322, USA}  

\begin{abstract} 
Using a simple, annealed model, some of the key features of the
recently introduced extremal optimization heuristic are
demonstrated. In particular, it is shown that the dynamics of local
search possesses a generic critical point under the variation of its
sole parameter, separating phases of too greedy (non-ergodic, jammed)
and too random (ergodic) exploration. Comparison of various local
search methods within this model suggests that the existence of the
critical point is essential for the optimal performance of the
heuristic.
\hfil\break PACS number(s): 02.60.Pn, 05.40.-a, 64.60.Cn,
75.10.Nr.
\end{abstract} 
\maketitle

\section{Introduction}
\label{introduction}
Many situations in physics and beyond require the solution of NP-hard
optimization problems, for which the typical time needed to ascertain
the exact solution apparently grows faster than any power of the
system size~\cite{G+J}. Examples in the sciences are the determination
of ground states for disordered
magnets~\cite{Dagstuhl04,Pal,Hartmann,P+Y,Houdayer99,eo_prl} or of
optimal arrangements of atoms in a compound~\cite{B+S} or a
polymer~\cite{Erzan,Frauenkorn,Prellberg}. With the advent of ever
faster computers, the exact study of such problems has become
feasible~\cite{P+A,Rinaldi}. Yet, with typically exponential
complexity of these problems, many questions regarding those systems
still are only accessible via approximate, heuristic
methods~\cite{Rayward}. Heuristics trade off the certainty of an exact
result against finding optimal or near-optimal solutions with high
probability in polynomial time. Many of these heuristics have been
inspired by physical optimization processes, for instance, simulated
annealing~\cite{Science,Salamon02} or genetic algorithms~\cite{Holland}.

Extremal optimization (EO) was proposed
recently~\cite{BoPe1,Dagstuhl04}, and has been used to treat a variety
of combinatorial~\cite{GECCO,BoPe2,BoPe3} and physical optimization
problems~\cite{eo_prl,Boettcher03b,EOSK}. Comparative studies with simulated
annealing~\cite{BoPe1,GECCO,EOperc} and other Metropolis~\cite{MRRTT} based
heuristics~\cite{D+S,Boettcher05b} have established EO as a successful
alternative for the study of NP-hard problems. EO has found a large
number of applications, for instance, in pattern
recognition~\cite{MB,Meshoul03}, signal filtering~\cite{YomTov},
transport problems~\cite{Sousa04a} molecular dynamics
simulations~\cite{Zhou05}, artificial
intelligence~\cite{Menai03a,Menai03b}, social
modeling~\cite{Duch05}, and $3d$ spin glass
models~\cite{D+S,Wang03,Onody03}. There are also a number of studies that
have explored basic features of the
algorithm~\cite{eo_jam,Boettcher05b},
extensions~\cite{Middleton04,Sousa03b,Sousa04b}, and rigorous
performance properties~\cite{Heilmann04,Hoffmann04}.

In this article, we will use a simple, annealed model of a generic
combinatorial optimization problem, introduced in Ref.~\cite{eo_jam},
to compare {\it analytically} certain variations of local search with
EO, and of Metropolis algorithms such as Simulated Annealing
(SA)~\cite{Science,Salamon02}. This comparison affirms the notion of
``optimization at the ergodic edge'' that motivated the $\tau$-EO
implementation~\cite{BoPe1,eo_prl}. This implementation possesses a
{\it single} tunable parameter, $\tau$, which separates a phase of
``greedy'' search from a phase of wide fluctuations, combining both
features at the phase transition into an ideal search heuristic for
rugged energy landscapes~\cite{Frauenfelder}. The model helps to
identify the distinct characteristics of different search heuristics
commonly observed in real optimization problems. In particular,
revisiting the model with a ``jammed'' state from Ref.~\cite{eo_jam}
proves the existence of the phase transition to be essential for the
superiority of EO, at least within a one-parameter class of local
search heuristics.  At the phase boundary, EO descends sufficiently
fast to the ground state with enough fluctuations to escape jams.

This article is organized as follows: In the next section, we
introduce the annealed optimization model, followed in
Sec.~\ref{search} by a short review of the local search heuristics
studied here, in particular, of one-parameter variations of EO and of
Metropolis-based search. Then, in Sec.~\ref{pedagogical} we compare
our analytical results for each heuristic in the annealed model. In
Sec.~\ref{theory} we show why versions of EO lacking a phase
transition fail to optimize well. We summarize our results and draw
some conclusions in Sec.~\ref{conclusion}.

\section{Annealed Optimization Model}
\label{evolution}
As described in Ref.~\cite{eo_jam}, we can abstract certain
combinatorial optimization problems into a simple, analytically
tractable model. To motivate this model we imagine a generic
optimization problem as consisting of a number of variables $1\leq
i\leq n$, each of which contributes an amount $-\lambda_i$ to the
overall cost per variable (or energy density) of the system,
\begin{eqnarray}
\epsilon=-\frac{1}{2n}\sum_{i=1}^n\lambda_i.
\label{defcosteq}
\label{lambdaeq}
\end{eqnarray}
(The factor $1/2$ arises because local cost are typically equally
shared between neighboring variables.) We call $\lambda_i\leq0$ the
``fitness'' of the variable, where larger values are better and
$\lambda_i=0$ is optimal for each variable. Correspondingly,
$\epsilon=0$ is the (optimal) ground state of the system. In a
realistic problem, variables are correlated such that not all of them
could be simultaneously of optimal fitness. But in our annealed model,
those correlations are neglected.

A concrete example for the above definitions is provided by a spin
glass with the Hamiltonian
\begin{eqnarray}
H=-\sum_{i,j}J_{i,j}\sigma_i\sigma_j
\label{Heq}
\end{eqnarray}
with some quenched random mix of bonds $J_{i,j}=\in\{-1,0,+1\}$ and
spin variables $\sigma_i=\pm1$~\cite{eo_prl}. With
$\lambda_i=\sigma_i\sum_jJ_{i,j}\sigma_j-\sum_j\vert J_{i,j}\vert$,
counting (minus) the number of violated bonds of each spin $i$ (among
its $\alpha_i$ non-zero bonds), it is $\epsilon=H/n+\epsilon_0$, where
$\epsilon_0$ is an insignificant constant.

We will consider that each variable $i$ is in one of $\alpha+1$
($\alpha_i=\alpha$ constant) different fitness states $\lambda_i$. We
can specify occupation numbers $n_a$, $0\leq a\leq\alpha$, for each
state $a$, and define occupation densities
$\rho_a=n_a/n~(a=0,\ldots,\alpha)$. Hence, any local search
procedure~\cite{Rayward} with single-variable updates, say, can be
cast simply as a set of evolution equations for the $\rho_a(t)$, i.~e.
\begin{eqnarray}
{\dot\rho}_b=\sum_a T_{b,a}Q_a.
\label{rhodoteq}
\end{eqnarray}
The $Q_a$ are the probabilities that a variable in state $a$ gets
updated; any local search process (based on updating a finite number
of variables) {\it defines} a unique set of $Q_a$, as we will see
below. The matrix $T_{b,a}$ specifies the net transition to state $b$
{\it given} that a variable in state $a$ is updated. This matrix
allows us to {\it design} arbitrary, albeit annealed, optimization
problems.  Both, ${\bf T}$ and ${\bf Q}$ generally depend on the
$\rho_a(t)$ as well as on $t$ explicitly.

We want to consider the different fitness states equally spaced, as in
the spin glass example above, where variables in state $a$ contribute
$a\Delta E$ to the energy to the system. Here $\Delta E>0$ is an
arbitrary energy scale. Thus minimizing the ``energy'' density
\begin{eqnarray}
\epsilon=\frac{1}{2}\sum_a a\rho_a\geq0,
\label{costeq}
\end{eqnarray} 
defines the optimization problem in this model.  Conservation of
probability and of variables implies the constraints
\begin{eqnarray}
\sum_a Q_a=1,&\qquad& \sum_a T_{a,b}=0\quad (0\leq b\leq\alpha),\\
\sum_a\rho_a(t)=1,&\qquad&\sum_a{\dot\rho}_a=0.
\label{normeq}
\end{eqnarray}

While this annealed model eliminates most of the relevant features of
a truly hard optimization problem, such as quenched randomness and
frustration~\cite{MPV}, two basic features of the evolution equations
in Eq.~(\ref{rhodoteq}) remain appealing: (1) The behavior of a system
with a large number of variables can be abstracted into a relatively
simple set of equations, describing their dynamics with a small set of
unknowns, and (2) the separation of update process, ${\bf T}$, and
update preference, ${\bf Q}$, lends itself to an analytical comparison
between different heuristics.

\section{Local Search Heuristics}
\label{search}
The annealed optimization model is quite generic for a class of
combinatorial optimization problems. But it was designed in particular
to analyze the ``Extremal Optimization'' (EO) heuristic~\cite{eo_jam},
which we will review next. Then we will present the update
probabilities ${\bf Q}$ through which each local search heuristic
enters into the annealed model in Sec.~\ref{evolution}. Finally, we
also specify the update probabilities ${\bf Q}$ for Metropolis-based
local searches, such as SA.

\subsection{Extremal Optimization Algorithm}
\label{eoalgorithm} 
Here we only give a quick review of the EO heuristic as we will use it
below. More substantive discussions of EO can be found
elsewhere~\cite{BoPe1,eo_prl,Dagstuhl04}.  EO is simply implemented
as follows: For a given configuration $\{\sigma_i\}_{i=1}^n$, assign
to each variable $\sigma_i$ an ``fitness''
\begin{eqnarray}
\lambda_i=-0,-1,-2,\ldots,-\alpha
\end{eqnarray}
(e.~g. $\lambda_i=-\{\#violated~bonds\}$ in the spin glass), so that
Eq.~(\ref{defcosteq}) is satisfied. Each variable falls into one of
only $\alpha+1$ possible states. Say, currently there are $n_{\alpha}$
variables with the worst fitness, $\lambda=-\alpha$, $n_{\alpha-1}$
with $\lambda=-(\alpha-1)$, and so on up to $n_0$ variables with the
best fitness $\lambda=0$. (Note that $n=\sum_in_i$.) Select an integer
$k$ ($1\leq k\leq n$) from some distribution, preferably with a bias
towards lower values of $k$. Determine $0\leq a\leq\alpha$ such that
$\sum_{b=a+1}^{\alpha}n_b<k\leq\sum_{b=a}^{\alpha}n_b$. Note that
lower values of $k$ would select a ``pool'' $n_a$ with larger value of
$a$, containing variables of lower fitness. Finally, select one of the
$n_a$ variables in state $a$ and update it {\em unconditionally.} As a
result, it and its neighboring variables change their fitness. After
all the effected $\lambda$'s and $n$'s are reevaluated, the next
variables is chosen for an update, and the process is repeated. The
process would continue to evolve, unless an extraneous stopping
condition is imposed, such as a fixed number of updates. The output of
local search with EO is the best configuration, with the lowest
$\epsilon$ in Eq.~(\ref{lambdaeq}), found up to the current update
step.

Clearly, a random selection of variables for such an update, without
further input of information, would not advance the local search
towards lower-cost states.  Thus, in the ``basic'' version of
EO~\cite{BoPe1}, each update one variable among those of worst fitness
would be made to change state (typically chosen at random, if there is
more than one such variable).

This provides a {\it parameter-free} local search of some
capability. But variants of this basic elimination-of-the-worst are
easily conceived. In particular, Ref.~\cite{BoPe1} already proposed
$\tau$-EO, a one-parameter ($\tau$) selection with a bias for
selecting variables of poor fitness on a slowly varying (power-law)
scale over the {\it ranking} $1\leq k\leq n$ of the variables by their
$\lambda_i$.  In detail, $\tau$-EO is characterized by a power-law
distribution over the fitness-ranks $k$,
\begin{eqnarray}
P_{\tau}(k)=\frac{\tau-1}{1-n^{1-\tau}} k^{-\tau}\quad(1\leq k\leq n).
\label{taueq}
\end{eqnarray}
It is a major point of this paper to demonstrate the usefulness of
this choice. Hence, we will compare the effect of this choice with a
plausible alternatives, $\mu$-EO, which uses an exponential scale,
\begin{eqnarray}
P_{\mu}(k)=\frac{e^\mu-1}{1-e^{-\mu n}} e^{-\mu k}\quad(1\leq k\leq
n).
\label{mueq}
\end{eqnarray}
In fact, we show that the exponential cut-off $\mu$ in $\mu$-EO, which
is fixed during a run, provides inferior results to $\tau$-EO. Unlike
$\tau$-EO, $\mu$-EO does not have a critical point affecting the
behavior of the local search.

Although Ref.~\cite{Heilmann04} has shown rigorously, that an optimal
choice is given by using a sharp threshold when selecting ranks, the
actual value of this threshold at any point in time is typically not
obvious (see also Ref.~\cite{Hoffmann04}). We will simulate a sharp
threshold $s$ ($1\leq s\leq n$) via
\begin{eqnarray}
P_{s}(k)\propto\frac{1}{1+e^{r(k-s)}} \quad(1\leq k\leq n)
\label{seq}
\end{eqnarray}
for $r\to\infty$. Since we can only consider fixed thresholds $s$,
which gives results similar in character to $\mu$-EO, it is not
apparent how to shape the rigorous results into a successful
algorithm.

\subsection{Update Probabilities for Extremal Optimization}
\label{eoupdates}
As described in Sec.~\ref{eoalgorithm} (and in Ref.~\cite{eo_jam}),
each update of $\tau$-EO a variable is selected based on its rank
according to the probability distribution in Eq.~(\ref{taueq}). When a
rank $k(\leq n)$ has been chosen, a variable is randomly picked from
state $\alpha$, if $k/n\leq\rho_{\alpha}$, from state $\alpha-1$, if
$\rho_{\alpha}<k/n\leq\rho_{\alpha}+\rho_{\alpha-1}$, and so on. We
introduce a new, continuous variable $x=k/n$, for large $n$
approximate sums by integrals, and rewrite $P(k)$ in Eq.~(\ref{taueq})
as
\begin{eqnarray} 
p(x)=\frac{\tau-1}{n^{\tau-1}-1} x^{-\tau}\quad\left(\frac{1}{n}\leq
x\leq 1\right),
\label{newtaueq}
\end{eqnarray}
where the maintenance of the low-$x$ cut-off at $1/n$ will turn out to
be crucial. Now, the average likelihood in EO that a variable in a
given state is updated is given by
\begin{eqnarray}
Q_{\alpha}&=&\int_{1/n}^{\rho_{\alpha}} p(x)dx=
\frac{1}{1-n^{\tau-1}}\left(\rho_{\alpha}^{1-\tau}-n^{\tau-1}\right),\nonumber\\
\medskip
Q_{\alpha-1}&=&\int_{\rho_{\alpha}}^{\rho_{\alpha}+\rho_{\alpha-1}}
p(x)dx=\frac{1}{1-n^{\tau-1}}\left[\left(\rho_{\alpha-1}+
\rho_{\alpha}\right)^{1-\tau}-\rho_{\alpha}^{1-\tau}\right],\nonumber\\
\medskip
&\ldots&\nonumber\\
\medskip
Q_0&=&\int_{1-\rho_0}^{1} p(x)dx=\frac{1}{1-n^{\tau-1}}
\left[1-\left(1-\rho_0\right)^{1-\tau}\right],
\label{qeq}
\end{eqnarray}
where in the last line the norm $\sum_a\rho_a=1$ was used. These
values of the $Q$'s completely describe the update preferences for
$\tau$-EO at arbitrary $\tau$.

Alternatively, if we consider the $\mu$-EO algorithm introduced in
Eq.~(\ref{mueq}), we have to replace the power-law distribution in
Eq.~(\ref{newtaueq}) with an exponential distribution:
\begin{eqnarray} 
p(x)=\frac{\mu}{1-e^{-\mu(1-1/n)}} e^{-\mu(x-1/n)}\quad\left(\frac{1}{n}
\leq x\leq 1\right),
\label{newmueq}
\end{eqnarray}
Hence, for $\mu$-EO we have
\begin{eqnarray}
Q_{\alpha}&=&\frac{1-e^{-\mu(\rho_\alpha-1/n)}}{1-e^{-\mu(1-1/n)}},\nonumber\\
\medskip
Q_{\alpha-1}&=&\frac{e^{-\mu(\rho_\alpha-1/n)}-e^{-\mu(\rho_\alpha+\rho_{\alpha-1}-1/n)}}{1-e^{-\mu(1-1/n)}},\nonumber\\
\medskip
&\ldots&\nonumber\\
\medskip
Q_0&=&\frac{e^{-\mu(1-\rho_0-1/n)}-e^{-\mu(1-1/n)}}{1-e^{-\mu(1-1/n)}}.
\label{muqeq}
\end{eqnarray}

Similarly, we can proceed with the threshold distribution in
Eq.~(\ref{seq}) to obtain
\begin{eqnarray}
p_{s}(x)\propto\frac{1}{1+e^{r(nx-s)}} \quad(\frac{1}{n}\leq x\leq 1),
\label{newseq}
\end{eqnarray}
with some proper normalization. While all the integrals to obtain
${\bf Q}$ are elementary, we do not display the rather lengthy results
here.

Note that all the update probabilities in each variant of EO are {\it
independent} of ${\bf T}$ (i.~e. any particular model), which remain
to be specified. This is quite special, as the following case of
Metropolis algorithms shows.

\subsection{Update Probabilities for Metropolis Algorithms}
\label{metropolis}
It is more difficult to construct ${\bf Q}$ for Metropolis-based
algorithms~\cite{MRRTT} like simulated
annealing~\cite{Science,Salamon02}. Let's assume that we consider a
variable in state $a$ for an update. Certainly, $Q_a$ would be
proportional to $\rho_a$, since variables are randomly selected for an
update. The Boltzmann factor $e^{-\beta\Delta E_a}$ for the potential
update from time $t\to t+1$ of a variable in $a$, aside from the
inverse temperature $\beta(t)$, only depends on the entries for
$T_{a,b}$:
\begin{eqnarray}
\Delta E_a&=&n\Delta \epsilon_a,\nonumber\\
\medskip
&=&\frac{n}{2}\left[\sum_bb\rho_b(t+1)-\sum_bb\rho_b(t)\right]_a,\nonumber\\
\medskip
&\sim&\frac{n}{2}\left[\sum_bb{\dot\rho}_b(t)\right]_a,\nonumber\\
\medskip
&=&\frac{n}{2}\left[\sum_bb\sum_cT_{b,c}Q_c \right]_a,\nonumber\\
\medskip
&=&\frac{n}{2}\sum_bbT_{b,a},
\end{eqnarray}
where the subscript $a$ expresses the fact that it is a {\it given}
that a variable in state $a$ is considered for an update. Hence, we
find for the average probability of an update of a variable in state
$a$
\begin{eqnarray}
Q_a=\frac{1}{\cal N}\rho_a{\rm min}\left\{1,\exp\left[-\beta
\frac{n}{2}\sum_bbT_{b,a}\right]\right\},
\label{SAeq}
\end{eqnarray}
where the norm ${\cal N}$ is determined via $\sum_aQ_a=1$. Unlike for
EO, the update probabilities for SA are model-specific, i.~e. depend
on ${\bf T}$.

\section{Comparison of Local Search Heuristics}
\label{pedagogical}

\begin{figure}
\begin{minipage}[t]{0.45\linewidth} 
\centering
\vspace{3.2in} \includegraphics{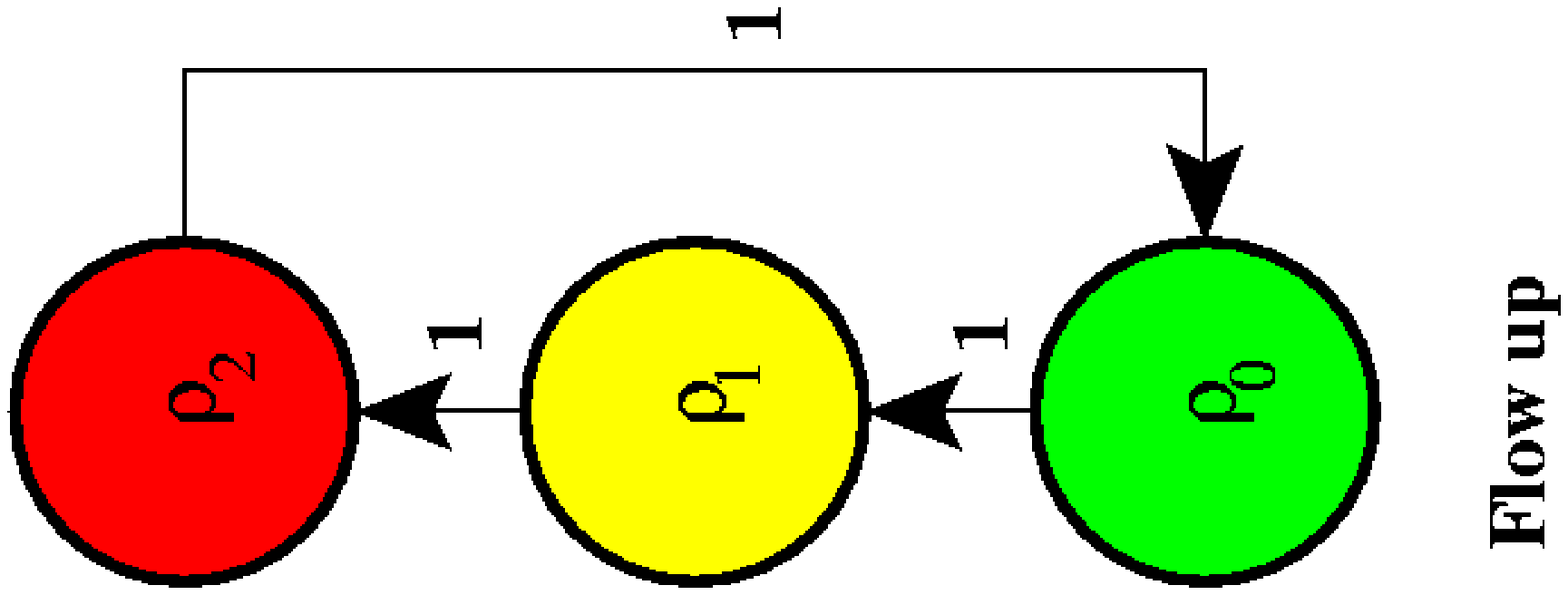}
\caption{Flow diagram with energetic barriers. Arrows indicate the
average number of variables transfered, $nT_{b,a}$, from a state $a$
to a state $b$, given that a variable in $a$ gets updated. Diagonal
elements $T_{a,a}$ correspondingly are negative, accounting for the
outflow. Note that variables transferring from $\rho_1$ to $\rho_0$
most first jump up in energy to $\rho_2$. }
\label{flowplot} 
\end{minipage}
\hspace{0.2in}
\begin{minipage}[t]{0.45\linewidth} 
\centering
\vspace{3.2in} \includegraphics{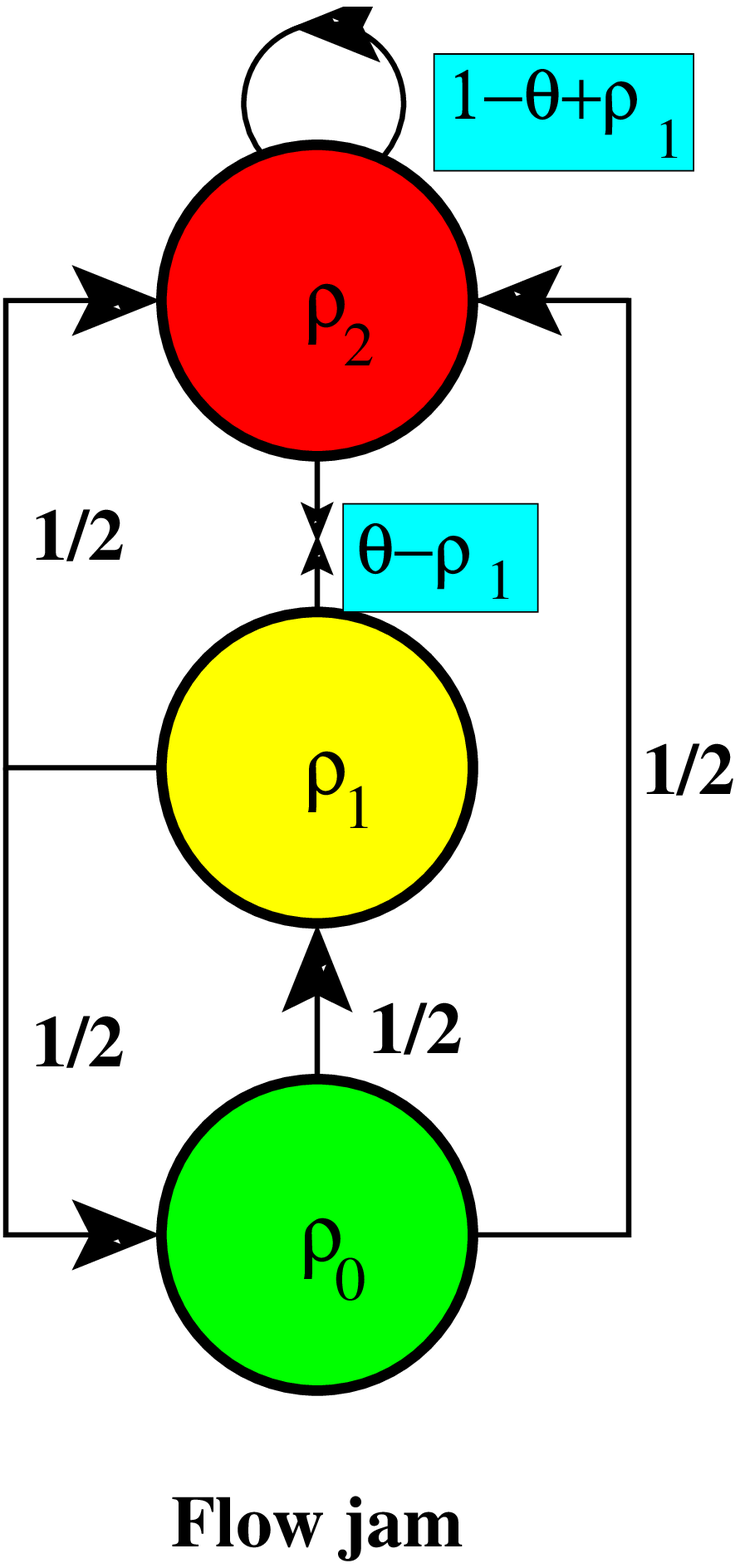}
\caption{Same as Fig.~\protect\ref{flowplot}, but with a model leading
to a jam. Variables can only transfer from $\rho_2$ to $\rho_0$
through $\rho_1$, but only if $\rho_1<\theta$. Once $\rho_1=\theta$,
flow down from $\rho_2$ ceases until $\rho_1$ reduces again.}
\label{jamflowplot}
\end{minipage}
\end{figure}

To demonstrate the use of these equations, we consider a simple model
of an energetic barrier with only three states $(\alpha=2)$ and a
constant flow matrix $T_{b,a}=[-\delta_{b,a}+\delta_{(2+b~{\rm
mod}~3),a}]/n$, depicted in Fig.~\ref{flowplot}. Here, variables in
$\rho_1$ can only reach their lowest-energy state in $\rho_0$ by first
jumping {\it up} in energy to $\rho_2$. Eq.~(\ref{rhodoteq}) gives
\begin{eqnarray}
{\dot\rho}_0=\frac{1}{n}\left(-Q_0+Q_2\right),\quad{\dot\rho}_1=\frac{1}{n}\left(Q_0-Q_1\right),\quad{\dot\rho}_2=\frac{1}{n}\left(Q_1-Q_2\right),
\label{flowupeq}
\end{eqnarray}
with ${\bf Q}$ discussed in Sec.~\ref{eoupdates} for the variants of
EO.

Given ${\bf T}$, we can now also determine the update probabilities
for Metropolis according to Eqs.~(\ref{SAeq}). Note that for $a=2$ we
can evaluate the $min$ as $1$, since $\sum_bbT_{b,a}<0$ always, while
for $a=0,1$ the $min$ always evaluates to the exponential. Properly
normalized, we obtain
\begin{eqnarray}
Q_0&=&\frac{\rho_0e^{-\beta/2}}{(1-e^{-\beta/2})\rho_2+e^{-\beta/2}},\quad
Q_1=\frac{\rho_1e^{-\beta/2}}{(1-e^{-\beta/2})\rho_2+e^{-\beta/2}},\nonumber\\
\medskip
Q_2&=&\frac{\rho_2}{(1-e^{-\beta/2})\rho_2+e^{-\beta/2}}.
\label{saqeq}
\end{eqnarray}

It is now very simple to obtain the stationary solution: For
${\dot{\bf\rho}}=0$, Eqs.~(\ref{flowupeq}) yields $Q_0=Q_1=Q_2=1/3$,
and we obtain from Eq.~(\ref{qeq}) for $\tau$-EO:
\begin{eqnarray}
\rho_0=1-\left(\frac{1}{3}n^{\tau-1}+\frac{2}{3}\right)^{\frac{1}{1-\tau}},\quad
\rho_2=\left(\frac{2}{3}n^{\tau-1}+\frac{1}{3}\right)^{\frac{1}{1-\tau}},\quad
\rho_1=1-\rho_0-\rho_2,
\label{eorhoeq}
\end{eqnarray}
for $\mu$-EO:
\begin{eqnarray}
\rho_0=\frac{1}{\mu}\ln\left[\frac{2}{3}+\frac{1}{3}e^{\mu\left(1-1/n\right)}\right],\quad
\rho_2=1-\frac{1}{\mu}\ln\left[\frac{1}{3}+\frac{2}{3}e^{\mu\left(1-1/n\right)}\right],\quad
\rho_1=1-\rho_0-\rho_2,
\label{mueorhoeq}
\end{eqnarray} 
and for Metropolis:
\begin{eqnarray}
\rho_0=\frac{1}{2+e^{-\beta/2}},\quad
\rho_1=\frac{1}{2+e^{-\beta/2}},\quad
\rho_2=\frac{e^{-\beta/2}}{2+e^{-\beta/2}}.
\label{sarhoeq}
\end{eqnarray} 
For EO with threshold updating, we obtain
\begin{eqnarray}
\rho_0&=&\frac{1}{3}-\frac{1}{3n}-\frac{s}{n}-\frac{1}{3nr}
\ln\left[1+e^{r(n-s)}\right]\nonumber\\
\medskip&&~~~~+\frac{1}{nr}
\ln\left[\left(e^{nr}+e^{rs}\right)\left(1+e^{r(1-s)}\right)^{\frac{1}{3}}+
e^{\frac{r}{3}(2n+1)}\left(1+e^{r(n-s)}\right)^{\frac{1}{3}}\right],\nonumber\\
\medskip
\rho_2&=&\frac{1}{3}+\frac{2}{3n}+\frac{s}{n}-\frac{2}{3nr}
\ln\left[1+e^{r(n-s)}\right]\nonumber\\
\medskip&&~~~~+\frac{1}{nr}\ln\left[
\left(e^{nr}+e^{rs}\right)\left(1+e^{r(1-s)}\right)^{\frac{2}{3}}+
e^{\frac{r}{3}(n+2)}\left(1+e^{r(n-s)}\right)^{\frac{2}{3}}\right],
\label{srhoeq}
\end{eqnarray} 
and, assuming a threshold anywhere between $1<s<n$, for $r\to\infty$:
\begin{eqnarray}
\rho_0=1-\frac{2s+1}{3n},\quad\rho_2=\frac{s+2}{3n},\quad\rho_1=1-\rho_0-\rho_2
\label{finalsrhoeq}
\end{eqnarray} 

Therefore, according to Eq.~(\ref{costeq}), Metropolis reaches its
best, albeit sub-optimal, cost $\epsilon=1/4>0$ at $\beta\to\infty$,
due to the energetic barrier faced by the variables in $\rho_1$, see
Fig.~\ref{flowplot}. (Since fluctuations from the mean are suppressed
in this model, even a slowly decreasing temperature schedule as in
Simulated Annealing would not improve results.) In turn, $\mu$-EO does
reach optimality ($\rho_0=1$, hence $\epsilon=0$), but only for
$\mu\to\infty$. Note that in this limit, $\mu$-EO reduces back to the
``basic'' version of EO discussed in Sec.~\ref{eoalgorithm}. The
result for threshold updating in EO are more promising: near-optimal
results are obtained, to within $O(1/n)$, for any finite threshold
$s$. But again, results are best for small $s\to1$, in which limit we
revert back to ``basic'' EO.

The result for $\tau$-EO is most remarkable: For $n\to\infty$ at
$\tau<1$ EO remains sub-optimal, but reaches the optimal cost {\it for
all} $\tau>1$! As discussed in Ref.~\cite{eo_jam}, this transition at
$\tau=1$ separates an (ergodic) random walk phase with too much
fluctuation, and a greedy descent phase with too little fluctuation,
which would trap $\tau$-EO in problems with broken ergodicity
\cite{Palmer}. This transition derives {\it generically} from the
scale-free power-law in Eq.~(\ref{taueq}), as was already argued on
the basis of numerical results for real NP-hard problems in
Refs.~\cite{BoPe1,BoPe2}.

\section{Jamming Model for $\mu$-EO}
\label{theory}
In this section, we revisit the ``jammed'' model treated in
Ref.~\cite{eo_jam} for $\tau$-EO and repeat that calculation for
$\mu$-EO. As in the example in Sec.~\ref{pedagogical}, $\mu$-EO proves
inferior to $\tau$-EO: Lacking the phase of optimal performance in the
$\tau$-parameter space, the required fine-tuning of $\mu$ does not
succeed in satisfying the conflicting constraints imposed on the
search.
 
Naturally, the range of phenomena found in a local search of NP-hard
problems is not limited to energetic barriers. After all, so far we
have only considered constant entries for $T_{b,a}$. Therefore, in our
next model we want to consider the case of ${\bf T}$ depending
linearly on the $\rho_i$ discussed in Ref.~\cite{eo_jam} for
$\tau$-EO. This model highlights significant differences between the
$\tau$-EO and the $\mu$-EO implementation.

From Fig.~\ref{jamflowplot}, we can read off ${\bf T}$ and obtain for
Eq.~(\ref{rhodoteq}):
\begin{eqnarray}
 {\dot\rho}_0&=&\frac{1}{n}\left[-Q_0+\frac{1}{2}Q_1\right],\nonumber\\
\medskip
 {\dot\rho}_1&=&\frac{1}{n}\left[\frac{1}{2}Q_0-Q_1+
(\theta-\rho_1)Q_2\right],
\label{thresheq}
\end{eqnarray}.
and ${\dot\rho}_2=-{\dot\rho}_0-{\dot\rho}_1$ from Eq.(\ref{normeq}).
Aside from the dependence of ${\bf T}$ on $\rho_1$, we have also
introduced the threshold parameter $\theta$. In fact, if
$\theta\geq1$, the model behaves effectively like the previous model,
and for $\theta\leq0$ there can be no flow from state $2$ to the lower
states at all. The interesting regime is the case $0<\theta<1$, where
further flow from state $2$ into state $1$ can be blocked for
increasing $\rho_1$, providing a negative feed-back to the system. In
effect, the model is capable of exhibiting a ``jam'' as observed in
many models of glassy dynamics~\cite{Jaeger,Ben-Naim,Ritort}, and
which is certainly an aspect of local search processes. Indeed, the
emergence of such a ``jam'' is characteristic of the low-temperature
properties of spin glasses and real optimization problems: After many
update steps most variables freeze into a near-perfect local
arrangement and resist further change, while a finite fraction remains
frustrated (temporarily in this model, permanently in real problems)
in a poor local arrangement~\cite{PSAA}. More and more of the frozen
variables have to be dislodged collectively to accommodate the
frustrated variables before the system as a whole can improve its
state. In this highly correlated state, frozen variables block the
progression of frustrated variables, and a jam emerges.

\begin{figure}
\vskip 5.0in \includegraphics{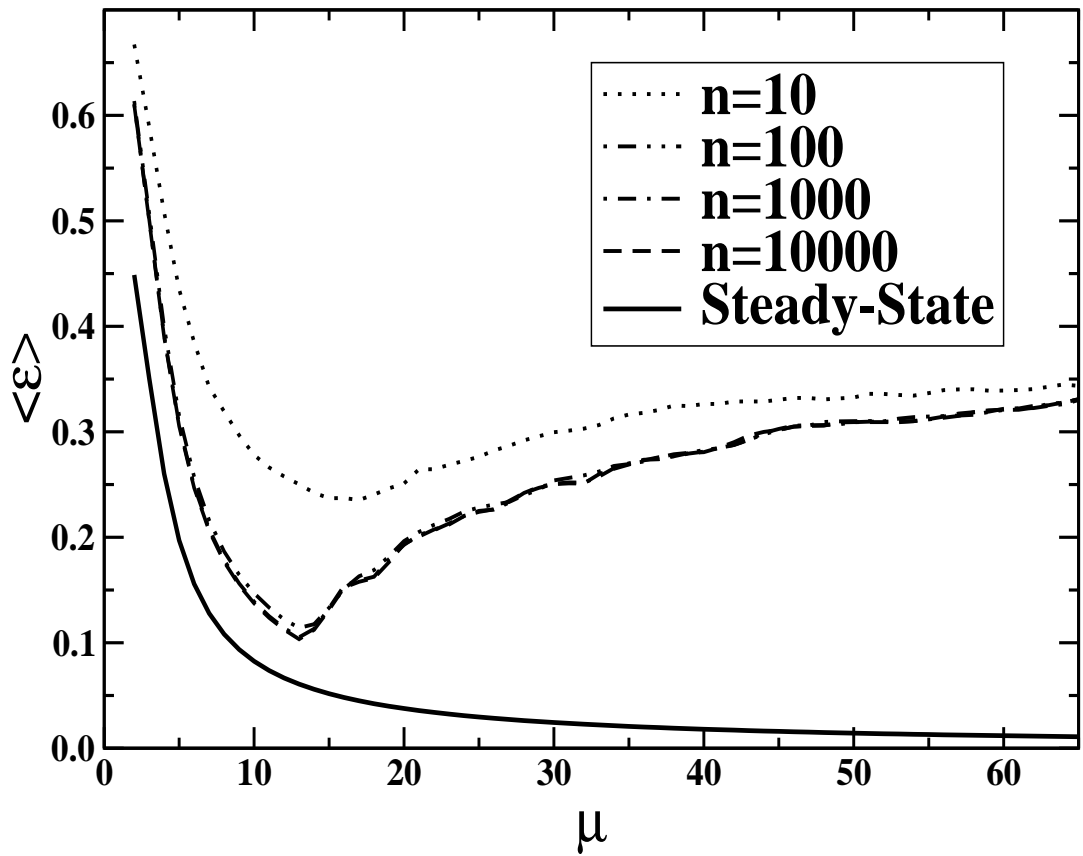}
\caption{Plot of the energy $\langle\epsilon\rangle$ averaged over
many $\mu$-EO runs after $t_{\rm max}=100n$ updates with different
initial conditions as a function of $\mu$ for $n=10$, 100, 1000, and
10000 and $\theta=1/2$. Plotted also is $\epsilon=\sum_aa\rho_a/2$ as
a function of $\mu$ resulting from the jam-free steady-state solution
($t\to\infty$) of Eqs.~(\protect\ref{eigeneq}-\protect\ref{sscosteq})
$n=\infty$. The plots show little variability with system size for
large $n$, and remain quite sub-optimal for finite $\mu$. As for
$\tau\to\infty$ in Ref.~\protect\cite{eo_jam},
$\langle\epsilon\rangle\to2/3-\theta/2-\theta^3/6=19/48\approx0.396$~\protect\cite{rem1}
for $\mu\to\infty$.}
\label{jammuplot}
\end{figure}

Inserting the set of Eqs.~(\ref{muqeq}) for $\alpha=2$ into the model
in Eqs.~(\ref{thresheq}), we obtain
\begin{eqnarray}
 {\dot\rho}_0&=&\frac{1}{n}\frac{1}{e^{\mu\left(1-1/n\right)}-1}
\left[1-\frac{3}{2}e^{\mu\rho_0}+
\frac{1}{2}e^{\mu\left(\rho_0+\rho_1\right)}\right],\nonumber\\
\medskip
{\dot\rho}_1&=&\frac{1}{n}\frac{1}{e^{\mu\left(1-1/n\right)}-1}
\left[-\frac{1}{2}+\frac{3}{2}e^{\mu\rho_0}-e^{\mu\left(\rho_0+\rho_1\right)}+
(\theta-\rho_1)\left(e^{\mu\left(1-1/n\right)}
-e^{\mu\left(\rho_0+\rho_1\right)}\right)\right],
\label{floweq}
\end{eqnarray}
At large times $t$, the steady state solution, ${\dot{\bf\rho}}=0$,
yields for $\rho_0$ after eliminating $\rho_1$ the implicit equation
\begin{eqnarray}
0=\frac{3}{2}-\frac{3}{2}e^{\mu\rho_0}+\left[\theta-\frac{1}{\mu}
\ln\left(3-2e^{-\mu\rho_0}\right)\right]\left(e^{\mu\left(1-1/n\right)}-
3e^{\mu\rho_0}\right),
\label{eigeneq}
\end{eqnarray}
and according to Eq.~(\ref{costeq}), again eliminating $\rho_1$ and
$\rho_2$ in favor of $\rho_0$, we can express the cost per variable as
\begin{eqnarray}
\epsilon&=&1-\rho_0-\frac{1}{2\mu}\ln\left(3-2e^{-\mu\rho_0}\right),\nonumber\\
\medskip
&\sim&\frac{1}{\mu}\ln\left[\sqrt{3}\left(1+\frac{1}{2\theta}\right)\right]
\quad(\mu\gg1),
\label{sscosteq}
\end{eqnarray}
Unlike the corresponding equations in Ref.~\cite{eo_jam}, which had a
phase transition similar to the solution for $\tau$-EO in
Sec.~\ref{pedagogical}, Eqs.~(\ref{eigeneq}-\ref{sscosteq}) have no
distinct features. In fact, as shown in Fig.~\ref{jammuplot},
$\epsilon(\mu)$ behaves similar to the solution for $\mu$-EO in
Sec.~\ref{pedagogical}: The relation is independent of $n$ to leading
order and only for $\mu\to\infty$, $\rho_0\to1$ and $\epsilon\to0$.

While the steady state ($t\to\infty$) features of this model do not
seem to be much different from the model in Sec.~\ref{pedagogical},
the dynamics at intermediate times $t$ is more subtle. In particular,
as was shown in Ref.~\cite{eo_jam}, a ``jam'' in the flow of variables
towards better fitness may ensue under certain circumstances. The
emergence of the jam depends on initial conditions, and its duration
will prove to get longer for larger values of $\mu$.  If the initial
conditions place a fraction $\rho_0>1-\theta$ already into the lowest
state, most likely no jam will emerge, since $\rho_1(t)<\theta$ for
all times, and the ground state is reached in $t=O(n)$ steps. But if
initially $\rho_1+\rho_2=1-\rho_0>\theta$, and $\mu$ is sufficiently
large, $\mu$-EO will drive the system to a situation where
$\rho_1\approx\theta$ by preferentially transferring variables from
$\rho_2$ to $\rho_1$. Then, further evolution becomes extremely slow,
delayed by the $\mu$-dependent, small probability that a variable in
state $1$ is updated ahead of all variables in state $2$.

Clearly, this jam is {\em not} a steady state solution of
Eq.~(\ref{floweq}). It is not even a meta-stable solution since there
are no energetic barriers. For instance, simulated annealing at zero
temperature would easily find the solution in $t=O(n)$ without
experiencing a jam. In reality, a hard problem would most certainly
contain combinations of jams, barriers, and possibly other
features.

To analyze the jam, we consider initial conditions leading to a jam,
$\rho_1(0)+\rho_2(0)>\theta$ and make the Ansatz
\begin{eqnarray}
\rho_1(t)=\theta-\eta(t)
\label{rho1eq}
\end{eqnarray}
with $\eta\ll1$ for $t\lesssim t_{\rm jam}$, where $t_{\rm jam}$ is
the time at which $\rho_0\to1$. To determine $t_{\rm jam}$, we apply
Eq.~(\ref{rho1eq}) to the evolution equations in (\ref{floweq}) to get
\begin{eqnarray}
 {\dot\rho}_0\sim\frac{1}{n}\frac{1}{e^{\mu\left(1-1/n\right)}-1}
\left[1-\frac{3}{2}e^{\mu\rho_0}+\frac{1}{2}e^{\mu\left(\rho_0+
\theta\right)}\right],
\label{rho0eq}
\end{eqnarray}
where the relation for ${\dot\rho}_1$ merely yields a self-consistent
equation to determine sub-leading corrections.

We can now integrate Eq.~(\ref{rho0eq}) from $t=0$ (assuming that any
jam emerges almost instantly) up to $t_{\rm jam}$, where $\rho_0=1$:
\begin{eqnarray}
t_{\rm jam}\sim n\left(e^\mu-1\right)\int_{\rho_0(0)}^{\rho_0(t_{\rm
jam})=1}
\frac{d\xi}{1-\left(\frac{3}{2}-\frac{1}{2}e^{\mu\theta}\right)e^{\mu\xi}}
\end{eqnarray}
The integral is easily evaluated, and we find for large values of
$\mu$:
\begin{eqnarray}
t_{\rm jam}\sim \frac{2n}{\mu}
e^{\mu(1-\theta-\rho_0(0))}\quad(\mu\gg1).
\label{tjameq}
\end{eqnarray}
Instead of repeating the lengthy calculation in Ref.~\cite{eo_jam} for
the ground state energy averaged over all possible initial conditions
for finite runtime $t_{\rm max}\propto n$, we can content ourselves
here with a few obvious remarks: A finite fraction of the initial
conditions will lead to a jam, hence will require a runtime $t_{\rm
max}\gg t_{\rm jam}$ to reach optimality. Yet, to reach a quality
minimum, say, $\epsilon\sim1/n$, would require $\mu\sim n\gg1$
according to Eq.~(\ref{sscosteq}). Thus, the require runtime to
resolve the jam would grow {\it exponentially} with system size $n$,
since from Eq.~(\ref{tjameq}) $t_{\rm jam}\sim e^{cn}$ with
$c=1-\theta-\rho_0(0)>0$, by definition of the jam above.

In conclusion, $\mu$-EO can never quite resolve the conflicting
demands of pursuing quality ground states with a strong bias for
selecting variables of low fitness (i.~e. $\mu\gg1$) and the ensuing
lack of fluctuations required to break out of a jam, which drives up
$t_{\rm jam}$. Simulations of this model with $\mu$-EO in
Fig.~\ref{jammuplot} indeed show that the best results for
$\langle\epsilon\rangle$ are obtained at intermediate values of $\mu$,
which converge to a large, constant error for increasing $n$. In
contrast, $\tau$-EO provides a range near $\tau_{\rm
opt}-1\sim1/\ln(n)$~\cite{eo_jam} with small enough $\tau$ to
fluctuate out of any jam in a time near-linear in $n$ while still
attaining optimal results as it does for {\it any} $\tau>1$, see
e.~g. Sec.~\ref{pedagogical}.

\section{Conclusion}
\label{conclusion}
We have presented a simple model to analyze the properties of local
search heuristics. The model with a simple energetic barrier
demonstrates the characteristics of a number of these heuristics,
whether athermal (EO and its variants) or thermal
(Metropolis)~\cite{Boettcher05b}. In particular, it plausibly
describes a number of real phenomena previously observed for $\tau$-EO
in a tractable way. Finally, in a more substantive comparison on a
model with jamming, the exponential distribution over fitnesses,
$\mu$-EO proves unable to overcome the conflicting constraints of
resolving the jam while finding good solutions. This is in stark
contrast with the identical calculation in Ref.~\cite{eo_jam} using a
scale-free approach with a power-law distribution over fitnesses in
$\tau$-EO. In this approach, a sharp phase transition emerges
generically between an expansive but unrefined exploration on one side
(``ergodic'' phase), and a greedy but easily trapped search on the
other (``non-ergodic'' phase), with optimal performance near the
transition.


\begin{thebibliography}{}

\bibitem{G+J} M. R. Garey and D. S. Johnson,  {\it Computers and
Intractability, A Guide to the Theory of NP-Completeness}
(W. H. Freeman, New York, 1979).

\bibitem{Dagstuhl04}
{\it New Optimization Algorithms in Physics,}
Eds. H. Rieger and A. K. Hartmann, (Springer, Berlin, 2004).

\bibitem{Pal} K. F. Pal,
Physica A {\bf 223}, 283-292 (1996), and  {\bf 233}, 60-66 (1996).

\bibitem{Hartmann} A. K. Hartmann,
Phys. Rev. B {\bf 59}, 3617-3623 (1999), and  Phys. Rev. E {\bf 60},
5135-5138 (1999).

\bibitem{P+Y} M. Palassini and A.~P. Young,
Phys. Rev. Lett. {\bf 85}, 3017 (2000).

\bibitem{Houdayer99} 
J. Houdayer and O. C. Martin,
Phys. Rev. Lett. {\bf 83}, 1030 (1999).

\bibitem{eo_prl} S. Boettcher and A. G. Percus,
Phys. Rev. Lett. {\bf 86}, 5211-5214 (2001).

\bibitem{B+S} K.~K. Bhattacharya and J.~P. Sethna,
Phys. Rev. E {\bf 57}, 2553 (1998).

\bibitem{Erzan} E. Tuzel and A. Erzan,
Phys. Rev. E {\bf 61}, R1040 (2000).

\bibitem{Frauenkorn}
H. Frauenkron, U. Bastolla, E. Gerstner, P. Grassberger, and W. Nadler,
Phys. Rev. Lett. {\bf 80}, 3149-3152 (1998).

\bibitem{Prellberg}
T. Prellberg and J. Krawczyk,
Phys. Rev. Lett. {\bf 92}, 120602 (2004).

\bibitem{P+A} 
R. G. Palmer and J. Adler,  
Int. J. Mod. Phys. C {\bf 10}, 667 (1999).

\bibitem{Rinaldi} C. Desimone, M. Diehl, M. J\"unger, P. Mutzel,
G. Reinelt, G. Rinaldi,
J. Stat. Phys. {\bf 80}, 487-496 (1995).

\bibitem{Rayward} 
{\em Modern Heuristic Search Methods,}
Eds. V. J. Rayward-Smith, I. H.  Osman, and C. R. Reeves (Wiley, New
York, 1996).

\bibitem{Science} S. Kirkpatrick, C. D. Gelatt, and M. P. Vecchi,
Science {\bf 220}, 671-680 (1983).

\bibitem{Salamon02}
P. Salamon, P. Sibani, and R. Frost,
{\it Facts, Conjectures, and Improvements for Simulated Annealing}
(Society for Industrial \& Applied Mathematics, 2002).

\bibitem{Holland} J.~Holland, {\em Adaptation in Natural and
Artificial Systems\/} (University of Michigan Press, Ann Arbor, 1975).

\bibitem{BoPe1}  S. Boettcher and A. G. Percus,
Artificial Intelligence {\bf 119}, 275-286 (2000).

\bibitem{GECCO} S. Boettcher and A. G. Percus,
in {\it GECCO-99: Proceedings of the Genetic and Evolutionary
Computation Conference} (Morgan Kaufmann, San Francisco, 1999),
825-832.

\bibitem{BoPe2}
S. Boettcher and A. G. Percus,
Physical Review E {\bf 64}, 026114 (2001).

\bibitem{BoPe3}
S. Boettcher and A. G. Percus,
Phys. Rev. E {\bf 69}, 066703 (2004).


\bibitem{Boettcher03b}
S. Boettcher,
Phys. Rev. B {\bf 67}, R060403 (2003).


\bibitem{EOSK}
S. Boettcher,
{\it Extremal Optimization for {S}herrington-{K}irkpatrick Spin
Glasses,}
Euro. Phys. J. B (in press), condmat/0407130.

\bibitem{EOperc} S. Boettcher,
J. Math. Phys. A {\bf 32}, 5201-5211 (1999).

\bibitem{MRRTT}
N.~Metropolis, A.W.~Rosenbluth, M.N.~Rosenbluth,
A.H.~Teller and E.~Teller,
{\em Equation of state calculations by fast computing machines,}
J.~Chem. Phys. {\bf 21} (1953) 1087--1092.

\bibitem{D+S} J. Dall and P. Sibani
Comp. Phys. Comm. {\bf 141}, 260-267 (2001).

\bibitem{Boettcher05b}
S. Boettcher and P. Sibani,
Euro. Phys. J. B {\bf 44}, 317-326 (2005).

\bibitem{MB}
S. Meshoul and M. Batouche,
Lecture Notes in Computer Science {\bf 2449}, 330-337 (2002).

\bibitem{Meshoul03}
S. Meshoul  and M. Batouche,
Int. J. Pattern Rec. and AI {\bf 17}, 1111-1126 (2003).

\bibitem{YomTov}
E. Yom-Tov, A. Grossman, and G. F. Inbar,
Biological Cybernatics {\bf 85}, 395-399 (2001).

\bibitem{Sousa04a}
F. L. Sousa, V. Vlassov and F. M. Ramos,
Heat Transf. Eng. {\bf 25}, 34-45 (2004).

\bibitem{Zhou05}
T. Zhou, W.-J. Bai, L.-J. Cheng, and B.-B. Wang,
Phys. Rev. E {\bf 72}, 016702 (2005).

\bibitem{Menai03a}
M. E. Menai and M. Batouche,
Lecture Notes in Computer Science {\bf 2718}, 592-603 (2003).

\bibitem{Menai03b}
M. E. Menai and M. Batouche,
in {\it Proceedings of the International Conference on Artificial
  Intelligence, IC-AI2003,} Eds. H. R. Arabnia et. al., 257-262 (2003).

\bibitem{Duch05}
J. Duch and A. Arenas,
{\it Community detection in complex networks using {E}xtremal {O}ptimization,}
cond-mat/0501368.

\bibitem{Wang03}
J.-S. Wang and Y. Okabe,
J. Phys. Soc. Jpn. {\bf 72}, 1380 (2003).

\bibitem{Onody03}
R. N. Onody and P. A. {de Castro},
Physica A {\bf 322}, 247-255 (2003).

\bibitem{eo_jam}
S. Boettcher and M. Grigni,
J. Phys. A. {\bf 35}, 1109 (2002).

\bibitem{Middleton04}
A. A. Middleton,
Phys. Rev. E {\bf 69}, 055701 (R) (2004).

\bibitem{Sousa03b}
F. L. de Sousa and V. Vlassov and F. M. Ramos,
Lecture Notes in Computer Science {\bf 2723}, 375-376 (2003).

\bibitem{Sousa04b}
F. L. de Sousa,  F. M. Ramos, R. L. Galski,  and I. Muraoka,
in {\it Recent Developments in Biologically Inspired Computing,}
Eds. L. N. De Castro and F. J. Von Zuben (Idea Group  Inc., 2004).

\bibitem{Heilmann04} 
F. Heilmann, K.-H. Hoffmann,  and P. Salamon,
Europhys. Lett. {\bf 66}, 305-310 (2004).

\bibitem{Hoffmann04}
 K.-H. Hoffmann, F. Heilmann,  and P. Salamon,
Phys. Rev. E {\bf 70}, 046704 (2004).

\bibitem{Frauenfelder}
{\it Landscape Paradigms in Physics and Biology,}
Ed. H. Frauenfelder (Elsevier, Amsterdam, 1997).

\bibitem{MPV} M. M\'ezard, G. Parisi, and M. A. Virasoro,
{\it Spin Glass Theory and Beyond,}
 (World  Scientific, Singapore, 1987).

\bibitem{Palmer}
F. T. Bantilan and R. G. Palmer,
J. Phys. F: Metal Phys. {\bf 11}, 261-266 (1981).

\bibitem{Jaeger} H.~M. Jaeger, S.~R. Nagel, R.~P. Behringer,
Rev.~Mod.~Phys {\bf 68} 1259-1273 (1996).

\bibitem{Ben-Naim} E. Ben-Naim, J.~B. Knight, E.~R. Nowak,
H.~M. Jaeger, and S.~R. Nagel,
Physica D {\bf 123}, 380-385 (1998).

\bibitem{Ritort} F. Ritort,
Phys.~Rev.~Lett. {\bf 75}, 1190-1193 (1995).

\bibitem{PSAA} R. G. Palmer, D. L. Stein, E. Abrahams, and
P. W. Anderson, Phys.~Rev.~Lett. {\bf 53}, 958 (1984).

\bibitem{rem1}
Ref.~\cite{eo_jam}
has in error in Eq.~(28): The general expression for the energy
$\epsilon$ in the integrand, $\sum_{i=0}^2i\rho_i/2=\rho_1/2+\rho_2$,
should be replaced by $\theta/2+\rho_2$ in the jam, which leads to
this value for $\langle\epsilon\rangle$ for large $\tau$ or $\mu$,
instead of $7/16\approx0.44$ quoted there.
\end{thebibliography}
\end{document}